\documentclass[aps,prl,superscriptaddress,reprint]{revtex4-2}

\usepackage{dcolumn}
\usepackage{bm}
\usepackage{float}
\usepackage{braket}
\usepackage{svg}
\usepackage{amsmath}
\usepackage{amssymb} 
\usepackage{graphicx}
\usepackage{caption}
\usepackage[font=small, justification=Justified]{caption}
\usepackage{subcaption}
\usepackage[export]{adjustbox}
\usepackage{hyperref}
\hypersetup{
    colorlinks=true,
    linkcolor=blue,
    citecolor=blue,
    urlcolor=blue,
    }
\usepackage{xcolor}

\begin{document}
\preprint{APS/123-QED}

\title{Geometric Antibunching and Directional Shaping of Photon Anticorrelations}

\author{Blas Durá-Azorín}
\affiliation{Departamento de Física Teórica de la Materia Condensada, Universidad Autónoma de Madrid, E-28049 Madrid, Spain}
\affiliation{Condensed Matter Physics Center (IFIMAC), Universidad Autónoma de Madrid, E-28049 Madrid, Spain.} 
\affiliation{Instituto de Qu\'imica F\'isica Blas Cabrera (IQF), CSIC, 28006 Madrid, Spain}
\author{Alejandro Manjavacas}
\email{a.manjavacas@csic.es}
\affiliation{Instituto de Qu\'imica F\'isica Blas Cabrera (IQF), CSIC, 28006 Madrid, Spain}
\author{Antonio I. Fernández-Domínguez}
\email{a.fernandez-dominguez@uam.es}
\affiliation{Departamento de Física Teórica de la Materia Condensada, Universidad Autónoma de Madrid, E-28049 Madrid, Spain}
\affiliation{Condensed Matter Physics Center (IFIMAC), Universidad Autónoma de Madrid, E-28049 Madrid, Spain.} 

\begin{abstract}
We investigate the directional characteristics of photon statistics in  dimers of quantum emitters. 
For their analysis, we construct a two-point second-order correlation function that allows us to find a new mechanism for photon anticorrelation, termed as \emph{geometric antibunching}. This phenomenon is completely agnostic to the quantum state of the emitters and emerges from quantum interference effects due to the indistinguishability of different two-photon optical pathways. Finally, we explore its occurrence in emitters placed in the vicinity of a flat substrate and a nanosphere, demonstrating its tunnability through the different material and geometric parameters of these structures.
\end{abstract}

\maketitle


Photon statistics are the cornerstone of quantum optics. Since the experiments of Hanbury-Brown and Twiss on the coherence of thermal sources~\cite{Hanbury}, or those by Kimble and co-workers on resonance fluorescence in atomic ensembles~\cite{Kimble}, many efforts have focused on their characterization in time domain. These experimental findings were accompanied by an intense theoretical activity, pioneered by Glauber, who introduced the celebrated normalized second-order correlation function, $g^{(2)}$~\cite{Glauber1963_PRL,Glauber1963_PR}. These joint efforts led to concepts such as photon bunching and antibunching~\cite{Loudon1976}, two-photon interference~\cite{HOM1987}, or photon blockade~\cite{Tian1992}, all rooted in the quantum nature of light. Subsequently, advances in photon generation and detection spurred much research aiming to refine the understanding of intensity correlations, beyond their temporal dependence. In particular, their filtering in frequency domain was shown to be extremely insightful already in the early stages of the field~\cite{Aspect1980}, and have been exploited as a powerful spectroscopic resource since then~\cite{Moreau2001,Dorfman2016}. Thus, on the theory side, the description of frequency-resolved $g^{(2)}$ has attracted much attention for decades~\cite{Cresser1987,Joosten2000,delValle2012, JENTLA24}.   

Contrary to their temporal and spectral attributes, the spatial dependence of intensity correlations has been greatly overlooked so far. Initial theoretical studies focused on the coherent emission by two distant sources in free space~\cite{Mandel,Richter,Skornia}, demonstrating a strong angular modulation in $g^{(2)}$ due to quantum interference. These results were generalized to quantum emitters (QEs) prepared in different pure states~\cite{liberal}, undergoing arbitrary emitter-emitter interactions~\cite{Ficek2002}, and placed in a random dielectric environment~\cite{Carminati2015,Canaguier}. Only very recently, experimental evidence of the spatial dependence of the second-order correlation function, i.e. $g^{(2)}=g^{(2)}(\bold{r},\bold{r}^\prime)$, was reported~\cite{Bunching?, collective}. In this Letter, inspired by late advances in the control of light emission by QEs coupled to nanophotonic systems~\cite{Novotny2011}, we explore the directional shaping of intensity correlations, rather than of light intensity itself. We present a comprehensive description of the spatial characteristics of photon anticorrelations in QE dimers. First, deploying a general expression for $g^{(2)}(\bold
{r},\bold
{r}^\prime)$ in terms of the electromagnetic dyadic Green's tensor~\cite{Novotny}, we inspect the case of free-standing QEs to find that interference-induced perfect antibunching takes place in the system, completely independent of its quantum state. We apply the same formalism to the most elemental optical device, a perfect mirror, to reveal the occurrence of geometric zeros in the second-order correlation function when the QEs are placed above it, and analyzing its directional characteristics and dependence on the different length scales involved. By substituting the perfect mirror with a substrate made of realistic materials, we unveil the universal character of this \emph{geometric antibunching}. Finally, placing the QEs in the near-field of metallic and dielectric nanospheres, we show that $g^{(2)}(\bold{r},\bold{r}^\prime)$ can be tailored through the selective excitation of the multipolar optical resonances they sustain.

The starting point of our theoretical study is the pair of QEs, which are described as two-level systems with natural frequency $\omega_0$, and as point-like sources, with dipole moment operator $\hat{\boldsymbol{\mu}}_{i}=\boldsymbol{\mu}_{i}(\hat{\sigma_{i}} + \hat{\sigma_{i}}^{\dagger})$. The QE lower operator, $\hat{\sigma_{i}}$ ($i=1,2$), satisfies the fermionic anticommutation relation $\{\hat{\sigma_{i}},\hat{\sigma}^{\dagger}_{i}\}=1$, and as both QEs are distinguishable, it also fulfills the commutation relations $[\hat{\sigma}_{i}, \hat{\sigma}_{j}]=0$, $[\hat{\sigma}_{i}, \hat{\sigma}^{\dagger}_{j}]=0$ for $i\neq j$. We neglect dephasing and non-radiative effects, assuming that the emitters present a radiative-limited lifetime, $\gamma_0^{-1}$. Tracing out the photonic degrees of freedom in the Macroscopic QED Hamiltonian~\cite{Dung, Asenjo2017, liberal}, the positive part of the electric field operator in an arbitrary dielectric environment is ${\hat{\bold
{E}}^{(+)}(\bold
{r})=\tfrac{\omega_{0}^{2}}{\varepsilon_{0}c^2}\sum_{i=1}^{2}\mathbf{G}(\bold
{r},\bold
{r}_{i},\omega_{0})\cdot\boldsymbol{\mu}_{i}\hat{\sigma}_{i}}$, 
where $\mathrm{\mathbf{G}}(\bold{r},\bold{r}_{i},\omega_{0})$ is the electromagnetic dyadic Green's tensor~\cite{Novotny}, and $\bold{r}_{i}$ the position of the $i$-th QE. In what follows, we write 
$\mathbf{G}(\bold{r},\bold{r}_i,\omega_0)\cdot\boldsymbol{\mu}_i=\mu\,
\psi(\bold{r},\bold{r}_{i})\bold{U}(\bold{r})$ (with $|\mu_{i}|=\mu$), to decompose the spatial dependence of the electric field operator into optical-path and polarization terms.

The two-point second-order correlation function at zero delay is defined as $G^{(2)}(\bold{r}, \bold{r}^\prime)=\langle:\hat{I}(\bold{r})\hat{I}(\bold{r}^\prime):\rangle$ \cite{Ficekbook}, where $:$ indicates normal ordering, and the intensity operator is $\hat{I}(\bold{r})=\hat{\bold{E}}^{(-)}(\bold{r})\cdot\hat{\bold{E}}^{(+)}(\bold{r})$. We introduce the operator
\begin{flalign}
\hat{E}_{2;\alpha,\beta}^{(+)}(\bold{r},\bold{r}^\prime)
=&\hat{E}_{\alpha}^{(+)}(\bold{r})\hat{E}_{\beta}^{(+)}(\bold{r}^\prime)=\nonumber \\
=&\left(\frac{\omega_{0}^{2}\mu}{\varepsilon_{0}c^2}\right)^2 U_{\alpha}(\bold{r})U_{\beta}(\bold{r}^\prime)\Psi(\bold{r}, \bold{r}^\prime; \bold{r}_{1}, \bold{r}_{2}) \hat{\sigma}_{1}\hat{\sigma}_{2},\label{E2}
\end{flalign}
which describes the two-point joint amplitude of electric field components $\alpha$ and $\beta$ at $\bold{r}$ and $\bold{r^\prime}$, respectively, and 
\begin{equation}
\Psi(\bold{r}, \bold{r}^\prime; \bold{r}_{1}, \bold{r}_{2})=\psi(\bold{r},\bold{r}_{1})\psi(\bold{r}^\prime,\bold{r}_{2}) + \psi(\bold{r}^\prime,\bold{r}_{1})\psi(\bold{r},\bold{r}_{2}) \label{Psi}
\end{equation}
appears as a result of the algebra of QE operators. This function can be interpreted as the optical-path dependence of the wavefuntion of the two-photon states generated by the QE dimer. We anticipate that this function is the main object of analysis in the following. With these definitions, the second-order correlation function can be written as $G^{(2)}(\bold{r}, \bold{r}^\prime)=\sum_{\alpha,\beta}\langle \hat{E}_{2;\alpha,\beta}^{(-)}(\bold{r},\bold{r}^\prime)\hat{E}_{2;\alpha,\beta}^{(+)}(\bold{r},\bold{r}^\prime)\rangle$ (note that we assume that the photon detectors at $\bold{r}$ and $\bold{r}^\prime$ are polarization agnostic). Evaluating the expectation value of the intensity operator, 
$\langle\hat{I}(\bold{r})\rangle=\left(\tfrac{\omega_{0}^{2}\mu}{\varepsilon_{0}c^2}\right)^2|\bold{U}(\bold{r})|^2\sum_{i, j=1}^{2}\psi^{*}(\bold{r},\bold{r}_{i})\psi(\bold{r},\bold{r}_{j})\langle \hat{\sigma}^{\dagger}_{i}\hat{\sigma}_{j}\rangle$,  we arrive at the expression for the normalized two-point second-order correlation function for the QE dimer 
\begin{widetext}
\begin{equation}
    g^{(2)}(\bold{r},\bold{r}^\prime) =\frac{G^{(2)}(\bold{r}, \bold{r}^\prime)}{\langle\hat{I}(\bold{r})\rangle\langle\hat{I}(\bold{r}^\prime)\rangle}=
    \frac{|\Psi(\bold{r}, \bold{r}^\prime; \bold{r}_{1}, \bold{r}_{2})|^{2}\langle \hat{\sigma}^{\dagger}_{1}\hat{\sigma}^{\dagger}_{2}\hat{\sigma}_{1}\hat{\sigma}_{2}\rangle}{\left(\sum_{i, j=1}^{2}\psi^{*}(\bold{r},\bold{r}_{i})\psi(\bold{r},\bold{r}_{j})\langle \hat{\sigma}^{\dagger}_{i}\hat{\sigma}_{j}\rangle\right)\left(\sum_{i, j=1}^{2}\psi^{*}(\bold{r}^\prime,\bold{r}_{i})\psi(\bold{r}^\prime,\bold{r}_{j})\langle \hat{\sigma}^{\dagger}_{i}\hat{\sigma}_{j}\rangle\right)},
\label{g2}
\end{equation}
\end{widetext}
where the $\bold{U}(\cdot)$ functions appearing in the numerator and denominator cancel out, rendering $g^{(2)}$ polarization-independent.

Equation~\eqref{g2} reveals that photon correlations in the emission of the QE dimer are not only determined by their quantum state, but also by the quantum interference effects between the different optical paths connecting QEs and detectors. In particular, $g^{(2)}(\bold{r},\bold{r}^\prime)$ vanishes if $\Psi(\bold{r}, \bold{r}^\prime; \bold{r}_{1}, \bold{r}_{2})=0$, independently of the correlator $\langle \hat{\sigma}^{\dagger}_{1}\hat{\sigma}^{\dagger}_{2}\hat{\sigma}_{1}\hat{\sigma}_{2}\rangle$, which encodes the quantum state of the system (and therefore its dependence on pumping, emitter interactions, and decoherence effects). This mechanism for intensity anticorrelation, governed by Equation~\eqref{Psi} and here termed as geometric antibunching, is inherently different from photon blockade or unconventional antibunchng~\cite{Zubizarreta2020}. Its physical origin can be traced back to Equations~\eqref{E2} and \eqref{Psi}, where the two-point joint amplitude operator itself nullifies when there is a perfectly destructive interference between the probability amplitudes, $|\psi(\bold{r}, \bold{r}_{1})\psi(\bold{r}^\prime, \bold{r}_{2})|$ and $|\psi(\bold{r}^\prime, \bold{r}_{1})\psi(\bold{r}, \bold{r}_{2})|$, which requires a minimum predictability for the system (difference of the two amplitudes above), in a similar way as in the which-way information problem~\cite{predictability}. 

\begin{figure}[b!]
\centering
    \includegraphics[width=0.95\columnwidth]{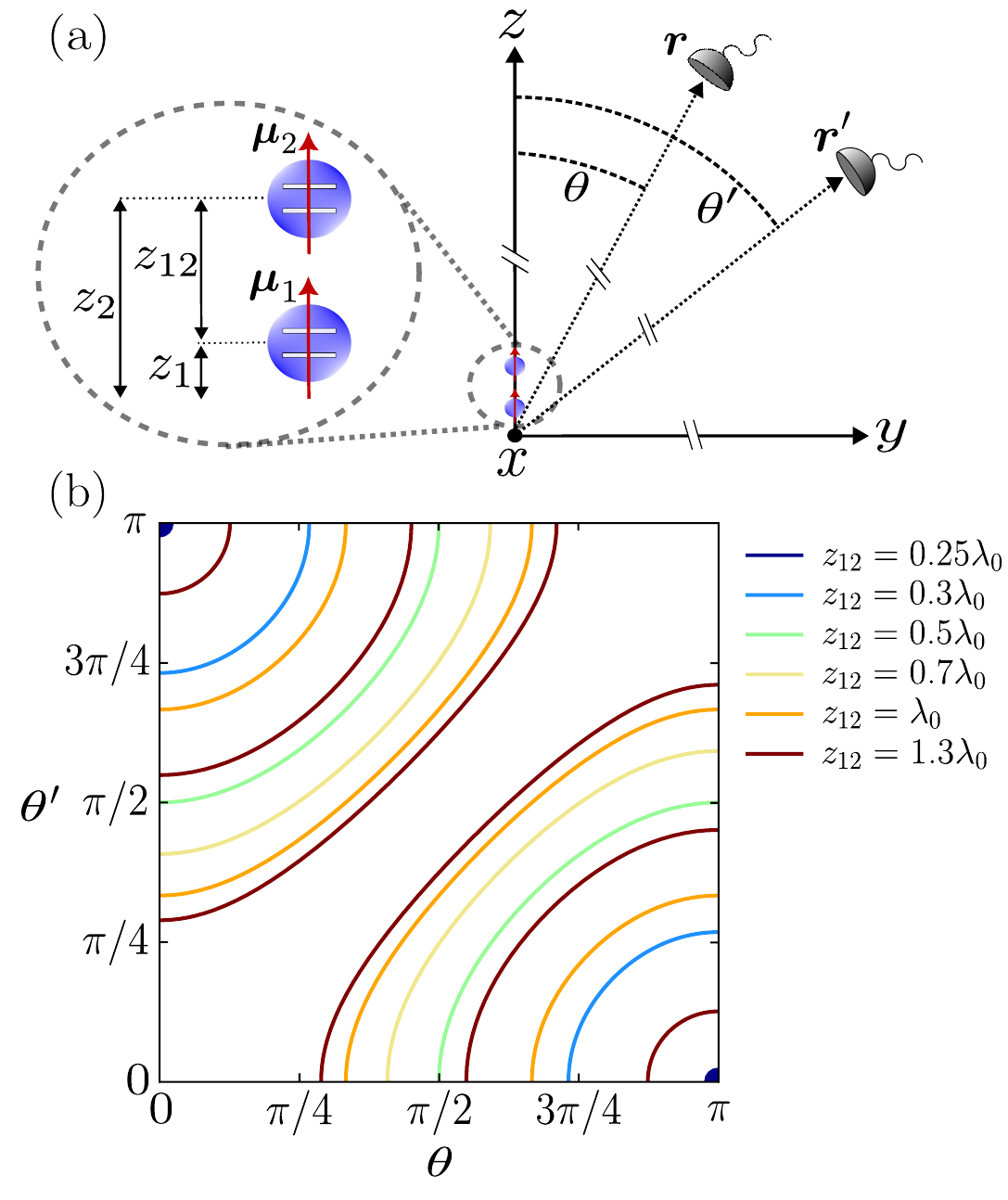}
    \caption{(a) Sketch of a QE dimer placed and oriented along the $z$-axis. The positions of the QEs are $z_{1}$ and $z_{2}$ and their distance $z_{12}$. Photon detectors are located in the far field, at polar angles $\theta$ and $\theta^\prime$. (b) Curves of geometric zeros in $g^{(2)}(\theta, \theta^\prime)$ in free space as a function of the directions of detection and for different $z_{12}$.}
   \label{fig1}
\end{figure}

We analyze first the emergence of geometric zeros in the two-point second--order correlation function in free space. In the far-field approximation~\cite{Novotny}, the electric field operator can be decomposed with ${\bold{U}(\bold{r})=(e^{ik_{0}r}/4\pi r)[\hat{\bold{\mu}}-\hat{\bold{r}}(\hat{\bold{r}}\cdot\hat{\boldsymbol{\mu}})]}$ and ${\psi(\bold{r},\bold{r}_{i})=e^{-ik_0\hat{\bold{r}}\cdot{\bold{r}_{i}}}}$, where $k_0=\omega_0/c$, $\hat{\bold{r}}=\bold{r}/|\bold{r}|$ and $\hat{\boldsymbol{\mu}}=\boldsymbol{\mu}/|\boldsymbol{\mu}|$. For these wavefunctions, the zero predictability constraint is satisfied for all $\hat{\bold{r}}$, and the only condition remaining for perfect geometric antibunching is that the phase difference between the two terms in Equation~\eqref{Psi} is an integer multiple of $\pi$. Figure~\ref{fig1}(a) shows a sketch of the system considered, with the two QEs located at, and aligned along $z$-direction. This highly symmetric configuration allows encoding all the directional dependence in terms of the polar angles $\theta$ and $\theta^\prime$. In Figure~\ref{fig1}(b), we plot the curves $g^{(2)}(\theta,\theta^\prime)=0$ for different inter-emitter distances,  $z_{12}$, normalized to the QE natural wavelength, $\lambda_0=2\pi/k_0$. Photon detection events along pairs of directions in the colored curves are perfectly anticorrelated, independently of the state of the QEs.  Figure~\ref{fig1}(b) reveals that there is a minimum inter-emitter distance for the occurrence of perfect geometric antibunching. At $z_{12}=\lambda_0/4$, the curves reduce to single points at $\{0, \pi\}$ and $\{\pi, 0\}$, for which the emitted intensity vanishes too and Equation~\eqref{g2} does not yield a zero in the second--order correlation function~\cite{Ficek2002}.\

We focus next on the emergence of geometric antibunching in QE dimers placed above a material substrate. We restrict our investigation to the axial configuration in Figure~\ref{fig1}(a), now in the presence of an underlying surface. For subwavelength $z_i$, we can express the far-field emission profile of each QE in the upper half-space as the sum of the contribution of its real dipole moment and that of its image dipole, located at $-z_{i}$ within the substrate and whose magnitude is determined by the Fresnel reflection coefficient, $r_{p}(\theta)$~\cite{Novotny}. As described in the Appendix, this spatial dependence is imprinted into  $\Psi$, modifying greatly the phenomenology of two-photon pathway interference taking place in free space.

\begin{figure}[h!]
\centering
    \includegraphics[width=0.95\columnwidth]{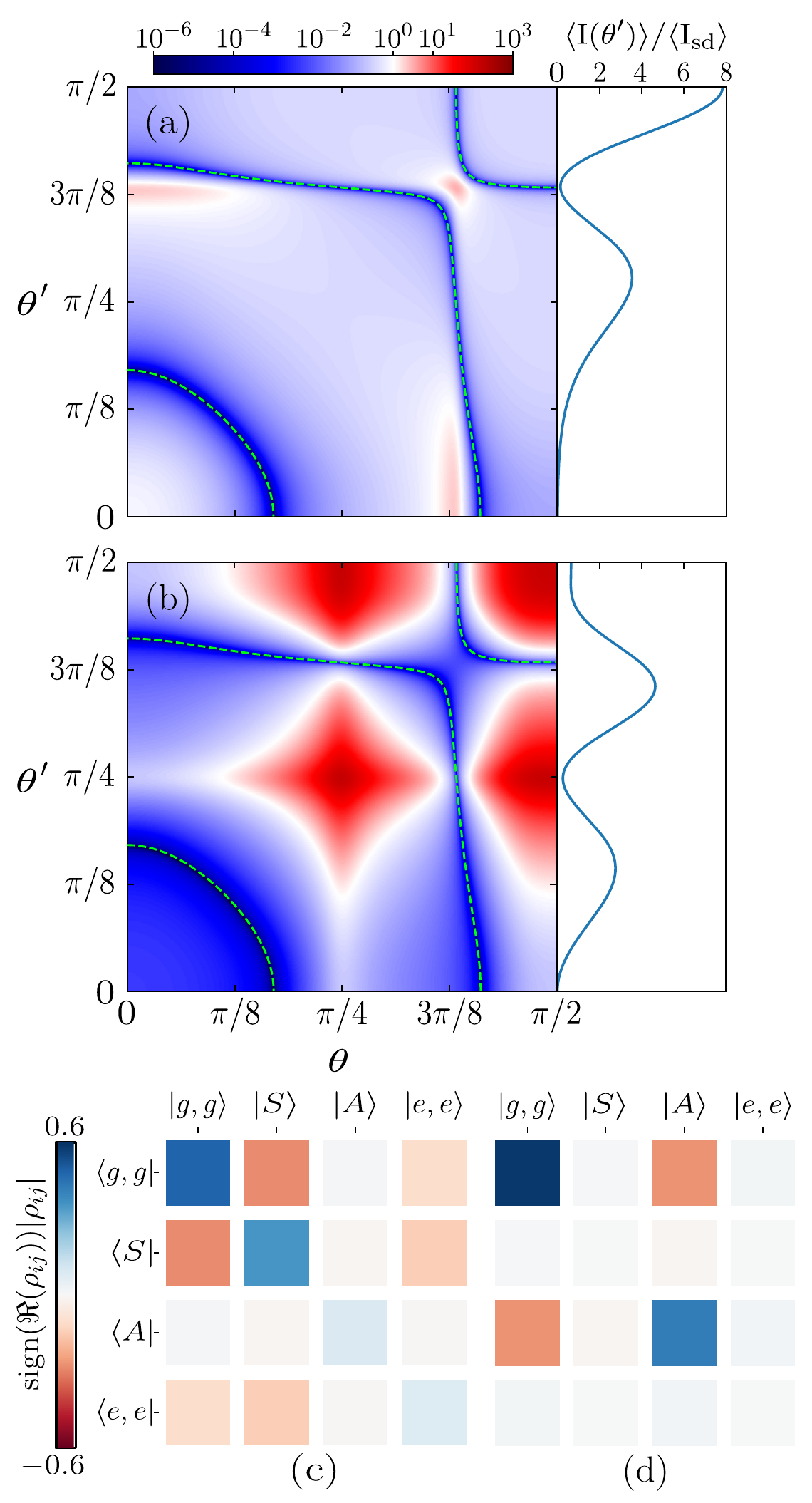}
    \caption{Maps of directional correlations, $g^{(2)}(\theta,\theta^\prime)$, for a vertically oriented QE dimer placed above a perfect mirror ($z_{1}=0.6\lambda_{0}$ and $z_{2}=0.8\lambda_{0}$). The laser drivings are set to target the Bell symmetric (a) and antisymmetric (b) states of the QE dimer. The lateral panels show directional intensity plots in both scenarios, normalized to $\langle I_{\rm sd}\rangle$ (the emission of a single emitter in free space at $\theta^\prime=\pi/2$). Panels (c) and (d) display quantum tomography maps of the states in panels (a) and (b).}
   \label{fig2}
\end{figure}

Figure~\ref{fig2} displays $g^{(2)}(\theta,\theta^\prime)$ for QEs placed at $z_{1}=0.6\lambda_{0}$ and $z_{2}=0.8\lambda_{0}$ above a perfect mirror ($r_{p}\rightarrow1$). We calculate the steady-state density matrix for the Lindbladian master equation~\cite{Dung}, accounting for the photon-induced coherent and dissipative interactions between the emitters (dictated by the dyadic Green's tensor), for two different coherent driving configurations (see the Appendix for more details). In Figure~\ref{fig2}(a), the laser frequency, $\omega_{\rm L}=\omega_0+g_{12}$ (where $g_{12}=g_{21}$ is the QE-QE coherent interaction strength), is set at resonance with the symmetric Bell state, $\ket{S}=[\ket{g,e}+\ket{e,g}]/\sqrt{2}$, and a strong driving amplitude, $E_{\rm L}\,\mu=\hbar\gamma_0$, is chosen for both QEs. In Figure~\ref{fig2}(b), $\omega_{\rm L}=\omega_0-g_{12}$ and the laser fields at the QEs have an opposite phase and same weak amplitude, $|E_{\rm L}\,\mu|=0.1\hbar\gamma_0$, to target the antisymmetric Bell state, $\ket{A}=[\ket{g,e}-\ket{e,g}]/\sqrt{2}$. The resulting photon correlation maps present distinct directional dependencies and exchanged regions of positive and negative correlations. These features are a consequence of the different steady-state of the QE dimer, which also manifests in the directional intensity plots, $\langle I(\theta') \rangle$, displayed in the lateral panels. Remarkably, there are three curves of vanishing $g^{(2)}$ (indicated by green dashed curves) that perfectly coincide in both correlation maps. This is the fingerprint of the geometric antibunching, which is completely independent of the state of the system. Figures~\ref{fig2}(c) and (d) display quantum tomography maps of the system density matrix in panels (a), and (b), respectively, showing that, indeed, the QEs are driven into significantly different quantum states.

\begin{figure}[t!]
\centering
    \includegraphics[width=0.9\columnwidth]{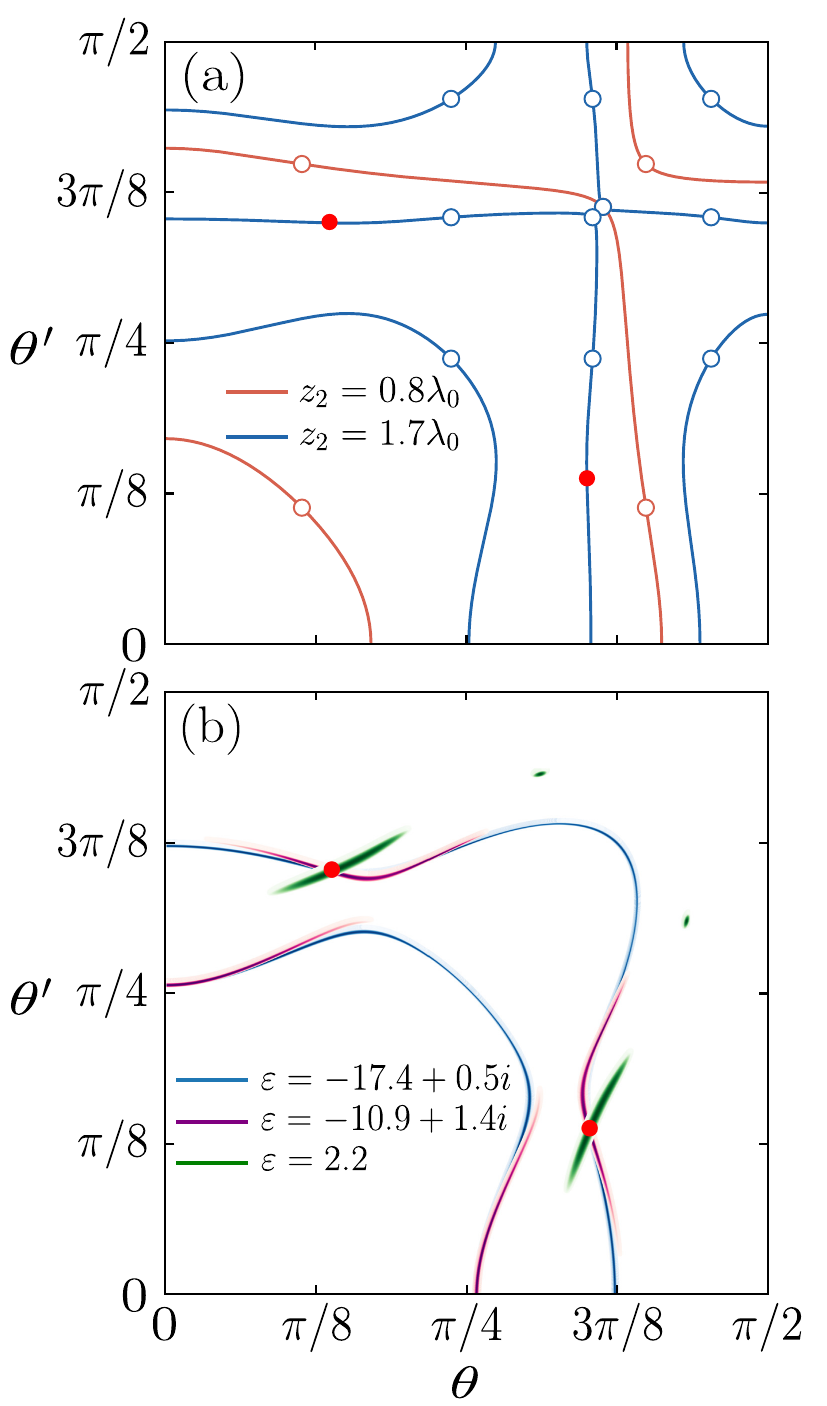}
    \caption{(a) Geometric zeros in the two-point second--order correlation function of a QE dimer placed above a perfect mirror, with $z_{1}=0.6\lambda_{0}$ and two different $z_2$ values. Empty dots correspond to the trivial correlation zeros caused by the quenching of one of the QEs. (b) Minima of $|\Psi|^{2}$ for substrates with three different permittivities, $\varepsilon$. In all cases, $z_{1}=0.6\lambda_{0}$ and $z_{2}=1.7\lambda_{0}$. The red dots in (a) and (b) indicate $\varepsilon$-independent zeros in $g^{(2)}(\theta,\theta^\prime)$.}
   \label{fig3}
\end{figure}

To gain insight into the geometric antibunching taking place in Figure~\ref{fig2}, we plot in Figure~\ref{fig3}(a) the vanishing condition for Equation~\eqref{Psi} within the $\theta\theta^\prime$-plane for $z_1=0.6\lambda_0$ and two different $z_2$ values: $0.8\lambda_0$ (red, the same case studied above) and $1.7\lambda_0$ (blue). The $\Psi=0$ curves in both cases present strong similarities (slightly displaced in the directional coordinates), although the number of branches increases with $z_2$. Contrary to free space, there is not a minimal inter-emitter distance for the vanishing of $g^{(2)}$ (note that $z_{12}=0.2\lambda_0$ for $z_2=0.8\lambda_0$). More importantly, there are several points along the diagonal $\theta=\theta^\prime$ at which there is perfectly antibunched emission. In these points, $\Psi(\bold{r}, \bold{r}^\prime; \bold{r}_{1}, \bold{r}_{2})=2\psi(\bold{\bold{r}}, \bold{r}_{1})\psi(\bold{\bold{r}}, \bold{r}_{2})$, which evidences that the correlations can only vanish through the complete quenching of one of the QEs. All of these trivial zeros, whose number also increases with $z_2$, are marked with empty dots in Figure~\ref{fig3}(a). 

We extend our investigation to material substrates different from a perfect mirror. We consider realistic values of the permittivity, $\varepsilon$, which present real and imaginary parts. This makes the Fresnel coefficients complex, altering both the amplitude and phase of the image dipole contributions to $\Psi(\bold{r}, \bold{r^\prime}; \bold{r}_{1}, \bold{r}_{2})$. Consequently, this function also becomes complex and, since the zeros of $\Re(\Psi)$ and $\Im(\Psi)$ do not coincide, the geometric zero conditions are not strictly fulfilled in general. Nevertheless, geometric antibunching still takes place for detection directions yielding a strong reduction of the numerator of Equation~\eqref{g2}. Figure~\ref{fig3}(b) displays the geometric minima in $g^{(2)}(\theta,\theta^\prime)$ for the same $z_{1}$ and $z_{2}$ as the blue curve of panel (a), and three different materials: two metals (blue, violet) and a dielectric (green). We use a threshold $|\Psi|^{2}=10^{-2}$ in white color, with smaller values indicated by darker colors. We observe that, for the two metals, the minima extend across the entire $\theta\theta^\prime$-plane and present strong similarities with their perfect mirror counterpart studied in panel (a). On the contrary, the minima for the dielectric are restricted to two localized regions in the map. Importantly, these regions include the positions marked with the red dots, which correspond to exact zeros of Equation~\eqref{Psi} that are independent of the substrate permittivity. These zeros do not originate from the destructive interference of different two-photon optical pathways in $\Psi(\bold{r}, \bold{r^\prime}; \bold{r}_{1}, \bold{r}_{2})$, but rather from the nullifying of all of them (see the Appendix). As a result, they depend only on the inter-emitter distance and not on the dielectric environment (they occur in the perfect mirror and free-space cases as well). These $\varepsilon$-independent zeros appear for a discrete set of directions, as explained in the Appendix.

\begin{figure}[t!]
\centering
    \includegraphics[width=0.9\columnwidth]{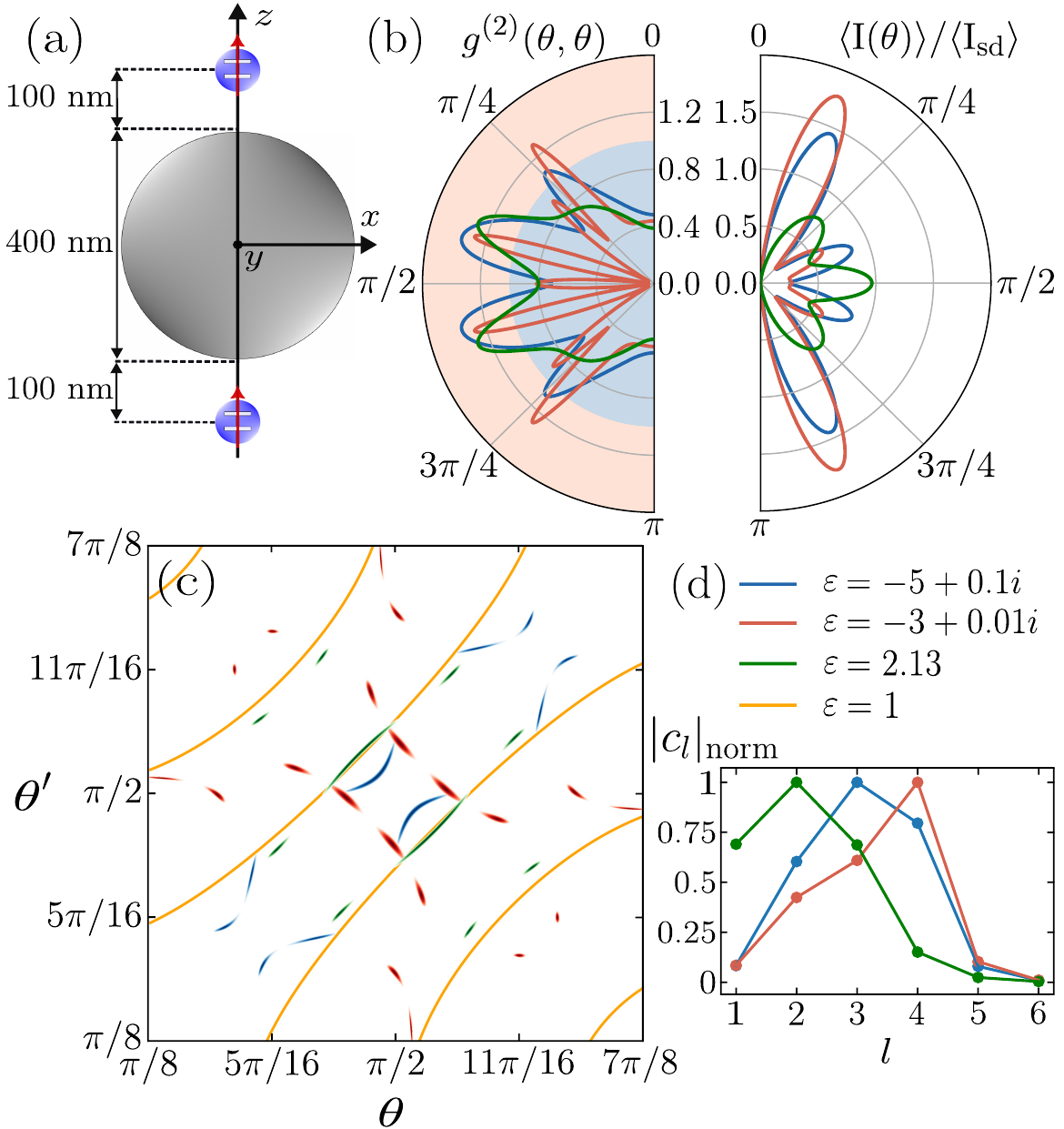}
    \caption{(a) Sketch of the QE dimer in the vicinity of a sphere of $200\,$nm radius. (b) Polar plots of $g^{(2)}(\theta,\theta)$ (left) and normalized $\langle I(\theta) \rangle$ (right) for three different $\varepsilon$. (c) Map of the minima of $|\Psi|^{2}$ for the three cases in (b), together with the exact geometric zeros in free space. We use the same threshold as in Figure~\ref{fig3}(b). (d) Normalized amplitude of the multipolar coefficients $c_{l}$.}
   \label{fig4}
\end{figure}

Finally, to illustrate the general character of geometric antibunching, we study the two-point second-order correlation function for a QE dimer placed in the vicinity of a sphere of $200\,$nm radius and different $\varepsilon$ (again, we consider two metals and a dielectric). Figure~\ref{fig4}(a) shows a sketch of the geometry, with the QEs placed above and below the nanosphere, at a distance of $100\,$nm from its surface. We compute the electromagnetic dyadic Green's tensor for this system using a quasi-analytical Mie expansion \cite{buhmann} described in the Appendix. Figure~\ref{fig4}(b) displays $g^{(2)}(\theta,\theta)$ and $\langle I(\theta) \rangle$ for a laser pumping scheme with $E_{\rm L}\,\mu=\hbar\gamma_0$ and $\hbar\omega_{\rm L}=\hbar\omega_0=3\,$eV. The polar plots of both magnitudes show a strong, but very different, directional dependence, which varies significantly from one permittivity to another. A more comprehensive perspective on $g^{(2)}(\theta,\theta^\prime)$ is offered by Figure~\ref{fig4}(c), which displays maps of geometric minima for the three permittivities under consideration, together with the results for free space ($\varepsilon=1$, in yellow). The rich directional phenomenology in these maps demonstrates that geometric correlations can be greatly tailored through the dielectric, near-field environment of the QEs.       

In this case, rather than relying on an image dipole picture, it is more insightful to interpret the geometric antibunching in Figure~\ref{fig4} in terms of quantum interference effects between the dipole of the QE dimer and the multipole contributions of the nanosphere to Equation~\eqref{Psi}. In Figure~\ref{fig4}(d), we plot the normalized amplitude of the coefficients corresponding to the different multipole terms of the electric field produced by the nanosphere, $c_{l}$, which we extract analytically. Notice that the three values of $\varepsilon$ under consideration result in different dominant multipoles. 
This allows us to analyze the impact of the multipole decomposition in $g^{(2)}(\theta,\theta^\prime)$. In particular, for the dielectric case (green), we observe a strong contribution of $l=1$. Therefore, the fields scattered by the nanosphere resemble those of a vertical dipole, and the curves of geometric zeros in Figure~\ref{fig4}(c) resemble those for $\varepsilon=1$. As the maximum of $|c_{l}|$ shifts towards larger $l$, the map of geometric antibunching becomes increasingly different from free space. It is also worth noting that there are not $\varepsilon$-independent geometric zeros in the nanosphere, as the lower symmetry of this structure prevents the concurrent, independent vanishing of all the contributions to $\Psi(\bold{r}, \bold{r}^\prime; \bold{r}_{1}, \bold{r}_{2})$.  

To conclude, we have investigated the directional nature of photon correlations in dimers of quantum emitters. First, we have constructed an appropriate theoretical tool for their analysis, the two-point second-order correlation function at zero delay. Applying it to free space, we have revealed a new mechanism for photon anticorrelations, termed geometric antibunching, which originates from interference among indistinguishable two-photon optical pathways and it is completely independent of the quantum state of the emitters. Next, by placing the dimer on top of a material substrate, we have studied how these far-field photon anticorrelations can be modified through the geometrical and material parameters of the system. Finally, we have focused on a more complex structure, a nanosphere, finding that the directionality of photon intensity and correlations can be tuned through the selective excitation of the different multipolar resonances it sustains. Our results open the way to new strategies to control and harness quantum light, by incorporating the spatial degrees of freedom in its theoretical description. We foresee efficient and versatile single-photon sources based on our findings, with application in quantum information and quantum sensing technologies.

\begin{acknowledgments}
This work has been generously supported by MCIN/AEI/10.13039/501100011033/FEDER under projects PID2021-126964OB-I00, TED2021-130552B-C21, and PID2022-137569NB-C42. BDA and AIFD also thank the support from the European Union's Horizon Program through grant 101070700.
\end{acknowledgments}

\onecolumngrid
\appendix
\section{Appendix}\label{ap}

\subsection{Lindblad master equation, decay rates and couplings}

The steady-state of the QE dimer is given by the following Lindblad master equation \cite{Dung}
\begin{equation}
    i[\rho, \hat{H}]+\sum_{i,j=1}^{2}\frac{\gamma_{ij}}{2}L_{\hat{\sigma}_{i},\hat{\sigma}_{j}}(\rho)=0,\label{master_eq}
\end{equation}
where $\rho$ is the steady-state density matrix of the system, $\hat{H}=-\hbar\Delta\sum_{i=1}^{2}\hat{\sigma}_{i}^{\dagger}\hat{\sigma}_{i} + (\hbar g_{12}\hat{\sigma}_{1}^{\dagger}\hat{\sigma}_{2} + \sum_{i=1}^{2}\Omega_{i}\hat{\sigma}_{i} + h.c.)$ is the Hamiltonian in the rotating frame of the driving laser (and within the RWA approximation), and $L_{\hat{\sigma}_{i},\hat{\sigma}_{j}}(\rho)=2\hat{\sigma}_{j}\rho\hat{\sigma}^{\dagger}_{i}- \{\hat{\sigma}^{\dagger}_{i}\hat{\sigma}_{j}, \rho\}$ are Lindbladian superoperators. The coefficients $\gamma_{ii}$ represent the decay rates of the QEs, while $\gamma_{12}$ and $g_{12}$ are their dissipative and coherent couplings, respectively. We assume a coherent driving produced by a laser with pumping rates $\Omega_{i} = E_{{\rm L}, i}\mu_i$ and detuning $\Delta=\omega_{\rm L}-\omega_{0}$, being $\omega_{\rm L}$ the frequency of the laser and $\omega_{0}$ the natural frequency of the QEs. We numerically solve this equation to find the density matrices used in Figure~\ref{fig2}(c-d), as well as to calculate $g^{(2)}(\theta,\theta^\prime)$ and $I(\theta)$ in Figure~\ref{fig2}(a-b) and Figure~\ref{fig4}(b). The parameters of the master equation can be written in terms of the electromagnetic dyadic Green's tensor of the QE environment~\cite{Dung}
\begin{equation}
    \gamma_{ij}=\frac{2\omega_{0}^{2}}{\hbar \varepsilon_{0}c^{2}}\boldsymbol{\mu}_{i}\cdot \mathrm{\Im\mathbf{G}}(\bold{r}_i,\bold{r}_j,\omega_{0})\cdot\boldsymbol{\mu}_{j},
\end{equation}
and
\begin{equation}
    g_{ij}=-\frac{\omega_{0}^{2}}{\hbar \varepsilon_{0}c^{2}}\boldsymbol{\mu}_{i
    }\cdot \mathrm{\Re\mathbf{G}}(\bold{r}_i,\bold{r}_j,\omega_{0})\cdot\boldsymbol{\mu}_{j}.
\end{equation}
To conclude, it is important to remark that we have not considered incoherent pumping and dephasing terms in the master equation. However, this simplification does not impact the principal result of our work---the emergence of geometrical zeros in $\Psi(\bold{r}, \bold{r}^{\prime}; \bold{r}_{1}, \bold{r}_{2})$---since these zeros are independent of the quantum state of the QE dimer.

\subsection{Quantum emitter dimer on a flat substrate}

The far-field operator of a QE located at $\bold{r}_{i}=z_{i}\boldsymbol{\hat{z}}$ (where $\boldsymbol{\hat{z}}$ is the normal to the flat substrate) with an arbitrarily oriented dipole moment is \cite{Novotny}
\begin{align}\nonumber
    \bold{\hat{E}^{(+)}}_{i}(\bold{r}) & =\begin{pmatrix}
                                    \hat{E}_{i,\theta}(\bold{r}) \\
                                    \hat{E}_{i,\phi}(\bold{r})
                                    \end{pmatrix}=\frac{\omega_{0}^{2}}{4\pi \varepsilon_{0} r c^2}e^{ik_{0}r} \\
                                    & \times  \begin{pmatrix}
                                    (\mu_{x} \cos\phi + \mu_{y} \sin\phi)\cos\theta(e^{-ik_{0}z_{i}\cos\theta} - r_{p}(\theta)e^{ik_{0}z_{i}\cos\theta})-\mu_{z}\sin\theta(e^{-ik_{0}z_{i}\cos\theta} + r_{p}(\theta)e^{ik_{0}z_{i}\cos\theta}) \\
                                     -(\mu_{x}\sin\phi - \mu_{y}\cos\phi)(e^{-ik_{0}z_{i}\cos\theta} + r_{s}(\theta)e^{ik_{0}z_{i}\cos\theta})
                                    \end{pmatrix}\hat{\sigma}_{i},
\label{field_substrate}
\end{align}
where $k_0=\omega/c$, $r_{p}(\theta)=(\varepsilon \mathrm \cos\theta - \sqrt{\varepsilon-\mathrm \sin^2\theta})/(\varepsilon \mathrm \cos\theta + \sqrt{\varepsilon-\mathrm \sin^2\theta})$ and $r_{s}(\theta)=(\mathrm \cos\theta - \sqrt{\varepsilon-\mathrm \sin^2\theta})/(\mathrm \cos\theta + \sqrt{\varepsilon-\mathrm \sin^2\theta})$ are the  Fresnel reflection coefficients for $p$ and $s$ polarization, respectively. In the limit of perfect mirror, this coefficients take the values $\varepsilon\rightarrow-\infty$ and $r_{p}\rightarrow1$, $r_{s}\rightarrow-1$. When the dipole moment of the QE is oriented along $\boldsymbol{\hat{z}}$, the electric field has only a $\theta$-component. In this case, the master equation parameters can be obtained analytically. 

The starting point of the calculation is the link between the QE decay rate and its emitted power, established by the Purcell factor, $\gamma/\gamma_{0}=P/P_{0}$~\cite{Novotny}. In this expression, $P=(\varepsilon_0 c/2)\int_{\partial S_{R}} |\bold{E}(\bold{R})|^2 R^2 d\Omega$ is the power radiated by a single QE over a sphere of radius $R\rightarrow\infty$ in a lossless environment, $\gamma$ is its radiative decay rate, whereas $P_{0}=\omega_{0}^{4}\mu^2/12\pi\varepsilon_{0}c^{3}$ and $\gamma_{0}=\omega_{0}^{3}\mu^{2}/3\hbar\varepsilon_{0}\pi c^{3}$ are their counterparts in free space~\cite{Novotny}. Using these quantities, we can easily derive the following relationship between the near- and far-field limits of the dyadic Green's tensor~\cite{Carminati2015}
\begin{equation}
    \hat{\boldsymbol{\mu}}\cdot \mathrm{\Im\mathbf{G}}(\bold{r}_{i},\bold{r}_{j},\omega_{0})\cdot\hat{\boldsymbol{\mu}}=k_{0}\int_{\partial S_{\infty}}\left(\hat{\boldsymbol{\mu}}\cdot\mathrm{\mathbf{G}_{ff}}^{\dagger}(\bold{r},\bold{r}_{i},\omega_{0})\cdot\mathrm{\mathbf{G}_{ff}}(\bold{r},\bold{r}_{j},\omega_{0})\cdot\hat{\boldsymbol{\mu}}\right)r^2d\Omega,
\label{green_function_rel}
\end{equation}
where $\boldsymbol{\hat{\mu}}=\boldsymbol{\mu}/|\boldsymbol{\mu}|$ is the unitary vector along the dipole moment of the QEs (we assume their orientation is the same). As anticipated, Equation~\eqref{green_function_rel} establishes a relationship between the dyadic Green's function evaluated at two arbitrary positions $\bold{r}_{i}$ and $\bold{r}_{j}$ and a surface integral of its mahematically simpler, far-field form, $\mathbf{G}_{\rm ff}$. We exploit this relationship to calculate $\gamma_{ii}$ and $\gamma_{12}$ for the QEs placed above of a perfect mirror substrate
\begin{eqnarray}
    \gamma_{ii}&=&\gamma_{0}\left(1-3\frac{2k_{0}z_{i}\cos(2k_{0}z_{i})-\sin(2k_{0}z_{i})}{(2k_0z_{i})^3}\right), \\ 
    \gamma_{12}&=&\gamma_{0}\frac{3}{k_{0}^6(z_{2}^2-z_{1}^2)^3}\Big\{(z_{1}+z_{2})^3k_{0}^3\big(\sin[k_{0}(z_{2}-z_{1})]-k_{0}(z_{2}-z_{1})\cos[k_{0}(z_{2}-z_{1})]\big) \nonumber  \\ 
   &&-k_{0}^4(z_{2}-z_{1})^3(z_{1}+z_{2})\cos[k_{0}(z_{1}+z_{2})]+(z_{2}-z_{1})^{3}k_{0}^3\sin[k_{0}(z_{1}+z_{2})]\Big\}.
\label{gamma12_substrate}
\end{eqnarray}
Since the fields vanish inside the perfect mirror, the integral in Equation~\eqref{green_function_rel} is restricted to the upper half-plane. Once $\gamma_{12}$ is calculated, we can use the Kramers-Kronig relations to compute $g_{12}$. By doing so, we obtain
\begin{eqnarray}
    g_{12}&=&-\frac{\gamma_{0}}{2}\frac{3}{k_{0}^6(z_{2}^2-z_{1}^2)^3}\big\{ (z_{1}+z_{2})^3{\rm sign}(z_{2}-z_{1})k_{0}^3\big(k_0(z_{2}-z_{1})\sin[k_{0}(z_{2}-z_{1})]+\cos[k_{0}(z_{2}-z_{1})]\big)\nonumber \\
   &&+k_{0}(z_{2}-z_{1})^3(z_{1}+z_{2})\sin[k_{0}(z_{1}+z_{2})] + (z_{2}-z_{1})^{3}k_{0}^3\cos[k_{0}(z_{1}+z_{2})]\big\}.
\label{g12_substrate}
\end{eqnarray}

\begin{figure}[h!]
\centering
    \includegraphics[width=0.95\textwidth]{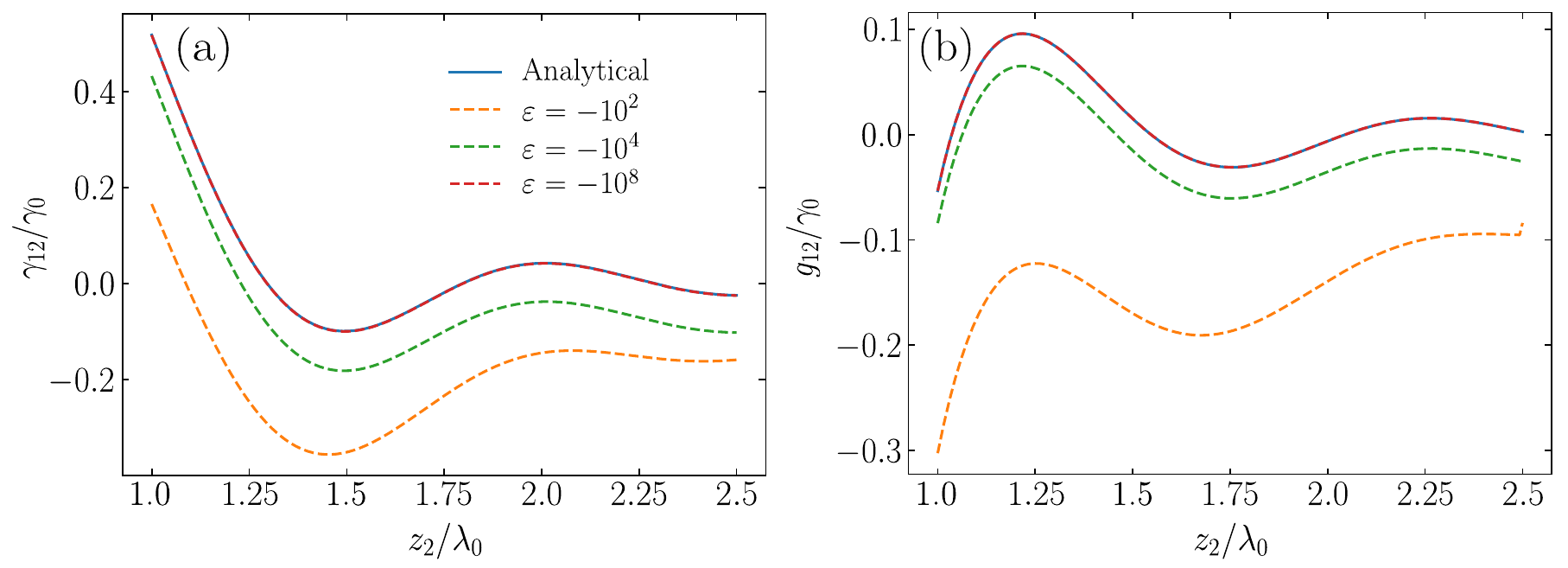}
    \caption{Dissipative (a) and coherent (b) coupling constants calculated for $z_1=0.6\lambda_0$ as a function of $z_2$. The solid curves display the results of the analytical expressions of Equations.~\eqref{gamma12_substrate} and \eqref{g12_substrate}, while the dashed curves correspond to the numerical evaluation of the dyadic Green's tensor for different values of the substrate permittivity $\varepsilon$, as indicated in the legend.} 
   \label{FIG_1_SM}
\end{figure}

Panels (a) and (b) of Figure~\ref{FIG_1_SM} show, respectively, the normalized dissipative and coherent coupling constants for $z_1=0.6\lambda_0$ and $z_2$ ranging from $\lambda_0$ to $2.5\lambda_0$, with $\lambda_0=2\pi/k_0$. The analytical predictions for the perfect mirror limit obtained from Equations.~\eqref{gamma12_substrate} and \eqref{g12_substrate} (solid curves) are compared against fully numerical calculations for various substrate permittivities (dashed curves). The latter are obtained using the $k$-space approach detailed in Ref.~\cite{Novotny}. We observe that the agreement between the numerical and analytical results improves for increasing $|\varepsilon|$ and the perfect mirror behaviour is fully reached for $\varepsilon=-10^8$. These panels demonstrate the validity of the expressions presented above for the master equation parameters in the case of a perfect mirror substrate.

We move next to the analysis of the optical path dependence of the two-photon wavefunction, $\Psi(\bold{r}, \bold{r}^\prime; \bold{r}_{1}, \bold{r}_{2})$. Noticing that, for QEs oriented along $\boldsymbol{\hat{z}}$,  Equation~\eqref{field_substrate} can be decomposed in terms of $U(\bold{r})=-\sin\theta (e^{ik_{0}r}/4\pi r)$ and $\psi(\theta,z_{i})=e^{i\alpha_{p}/2}(e^{-i(k_{0}z_{i}\cos\theta + \alpha_{p}/2)} + |r_{p}|e^{i(k_{0}z_{i}\cos\theta + \alpha_{p}/2)})$, the condition for the geometrical zeros in the second-order correlation function for a general $\varepsilon$ can be written as
\begin{align}\nonumber
    \Psi & =e^{i(\alpha_p+\alpha'_p)/2} \left[ e^{-i(k_{0}z_{1}\cos\theta + \alpha_{p}/2)}e^{-i(k_{0}z_{2}\cos\theta^\prime + \alpha_{p}^\prime/2)}+e^{-i(k_{0}z_{1}\cos\theta^\prime + \alpha^\prime_{p}/2)}e^{-i(k_{0}z_{2}\cos\theta + \alpha_{p}/2)} \right.\\ \nonumber
    & + |r_{p}^\prime|\left(e^{-i(k_{0}z_{1}\cos\theta + \alpha_{p}/2)}e^{i(k_{0}z_{2}\cos\theta^\prime + \alpha_{p}^\prime/2)}+e^{i(k_{0}z_{1}\cos\theta^\prime + \alpha^\prime_{p}/2)}e^{-i(k_{0}z_{2}\cos\theta + \alpha_{p}/2)} \right) \\ \nonumber
    & + |r_{p}|\left(e^{i(k_{0}z_{1}\cos\theta + \alpha_{p}/2)}e^{-i(k_{0}z_{2}\cos\theta^\prime + \alpha_{p}^\prime/2)}+e^{-i(k_{0}z_{1}\cos\theta^\prime + \alpha^\prime_{p}/2)}e^{i(k_{0}z_{2}\cos\theta + \alpha_{p}/2)} \right) \\ 
    & + \left. |r_{p}||r_{p}^\prime|\left(e^{i(k_{0}z_{1}\cos\theta + \alpha_{p}/2)}e^{i(k_{0}z_{2}cos\theta^\prime + \alpha_{p}^\prime/2)}+e^{i(k_{0}z_{1}\cos\theta^\prime + \alpha^\prime_{p}/2)}e^{i(k_{0}z_{2}\cos\theta + \alpha_{p}/2)} \right)\right]=0,
    \label{wavefunction_substrate}
\end{align}
where $r_p(\theta)=|r_p|e^{i\alpha_p}$ and $r_p(\theta^\prime)=|r_p^\prime|e^{i\alpha_p^\prime}$. The first term of the expression above is the same as in free space, and it accounts for the pathway interference experienced by the photons produced by the dipole moment of the QEs. The second and third terms correspond to the interference between the photons emitted by the dipoles of the QEs and their corresponding images embedded in the material substrate. Finally, the fourth term corresponds to the contribution of the image dipoles of both QEs. Note that, due to the high symmetry of the system, the first and fourth terms in Equation~\eqref{wavefunction_substrate} are complex conjugates of each other, except for the product of the Fresnel coefficient moduli, $|r_p||r_p^\prime|$. Similarly, the second and third ones are also complex conjugates, except for the ratio $|r_p|/|r_p^\prime|$. 

The geometrical zeros of $\Psi(\bold{r}, \bold{r}^{\prime}; \bold{r}_{1}, \bold{r}_{2})$  are given by all of the solutions of Equation~\eqref{wavefunction_substrate}. Now, we focus on a particular set of them that originate from the nullifying of each of the four terms of $\Psi$. As discussed above, only the first and second ones are independent, and hence, these zeros are insensitive to the Fresnel coefficients and the substrate permittivity. These are termed as $\varepsilon$-independent zeros and are given by $\cos \theta=n\lambda_{0}/2z_{12}$ and $\cos \theta^\prime=m\lambda_{0}/2z_{12}$ (with $n,m\in\mathbb{Z}$). These integers, however, are not arbitrary: the solutions must lie in the upper half-plane and must satisfy the inequalities $0<n,m<2z_{12}/\lambda_{0}$. Additionally, $\theta$ and $\theta^\prime$ cannot be identical and therefore, neither can $n$ and $m$. Consequently, the first two solutions are $\{n,m\}=\{1,2\}$ and $\{2,1\}$, which correspond to the red dots in Figure~\ref{fig3}. Furthermore, this implies that the inter-emitter distance $z_{12}$ required to achieve $\varepsilon$-independent zeros must exceed $\lambda_{0}$.

These substrate-agnostic zeros emerge even when the QE dipole moments are not oriented along $\boldsymbol{\hat{z}}$. In this case, the field cannot be decomposed as $\mathbf{G}(\bold{r},\bold{r}_i,\omega_0)\cdot\boldsymbol{\mu}=\mu\,
\psi(\bold{r},\bold{r}_{i})\bold{U}(\bold{r})$ because the functions $\psi$ are also polarization dependent. Nevertheless, we can write the components of Equation~\eqref{field_substrate} as
\begin{align}
 \hat{E}_{\theta}(\bold{r})&=\frac{\omega_{0}^{2}}{\varepsilon_{0}c^2}\left(U_{\theta, \parallel}(\boldsymbol{\mu}_{\parallel},\bold{r})\sum_{i=1}^2\psi_{\theta,\parallel}(\theta,z_{i})\hat{\sigma}_{i} + U_{\theta, z}(\mu_{z},\bold{r})\sum_{i=1}^2\psi_{\theta,z}(\theta,z_{i})\hat{\sigma}_{i}\right), \\
 \hat{E}_{\phi}(\bold{r})&=\frac{\omega_{0}^{2}}{\varepsilon_{0}c^2}U_{\phi, \parallel}(\boldsymbol{\mu}_{\parallel},\bold{r})\sum_{i=1}^2\psi_{\phi,\parallel}(\theta,z_{i})\hat{\sigma}_{i},
\end{align} 
 where
\begin{equation}
    U_{\theta, \parallel}(\boldsymbol{\mu}_{\parallel},\bold{r})=(e^{ik_{0}r}/4\pi r)[\mu_{x}\cos\phi + \mu_{y}\sin \phi]\cos \theta,
\label{U}
\end{equation}
\begin{equation}
    U_{\theta, z}(\mu_{z},\bold{r})=-(e^{ik_{0}r}/4\pi r)\mu_{z}\sin \phi,
\end{equation}
\begin{equation}
   U_{\phi, \parallel}(\boldsymbol{\mu}_{\parallel},\bold{r})=-(e^{ik_{0}r}/4\pi r)[\mu_{x}\sin \phi - \mu_{y}\cos \phi],
\end{equation}
and
\begin{equation}
    \psi_{\theta,\parallel}(\theta,z_{i})=e^{i\alpha_{p}/2}(e^{-i(k_{0}z_{i}\cos\theta + \alpha_{p}/2)} - |r_{p}|e^{i(k_{0}z_{i}\cos\theta + \alpha_{p}/2)}),
\end{equation}
\begin{equation}
    \psi_{\theta,z}(\theta,z_{i})=e^{i\alpha_{p}/2}(e^{-i(k_{0}z_{i}\cos\theta + \alpha_{p}/2)} + |r_{p}|e^{i(k_{0}z_{i}\cos\theta + \alpha_{p}/2)}),
\end{equation}
\begin{equation}
    \psi_{\phi,\parallel}(\theta,z_{i})=e^{i\alpha_{s}/2}(e^{-i(k_{0}z_{i}\cos\theta + \alpha_{s}/2)} + |r_{s}|e^{i(k_{0}z_{i}\cos\theta + \alpha_{s}/2)}).
\label{psi}
\end{equation}
When the probability amplitudes depend on polarization, geometric antibunching requires the nullifiying of the four two-point joint amplitude operators $\hat{E}_{2;\alpha,\beta}^{(+)}(\bold{r},\bold{r}^\prime)=\left(\frac{\omega_{0}^{2}\mu}{\varepsilon_{0}c^2}\right)^2 \Psi_{\alpha,\beta}(\bold{r}, \bold{r}^\prime; \bold{r}_{1}, \bold{r}_{2}) \hat{\sigma}_{1}\hat{\sigma}_{2}$, (with $\alpha,\beta=\theta,\phi$). Consequently, four polarization-dependent two-photon wavefunctions, $\Psi_{\alpha,\beta}$, must vanish. For example, the condition $\Psi_{\theta,\theta}=0$  yields
\begin{align}\nonumber
    \Psi_{\theta,\theta} = & U_{\theta, \parallel}(\boldsymbol{\mu}_{\parallel},\bold{r})U_{\theta, \parallel}(\boldsymbol{\mu}_{\parallel},\bold{r}^\prime)e^{i(\alpha_{p}+\alpha_{p}^\prime)/2}\left[\psi_{\theta,\parallel}(\theta,z_{1})\psi_{\theta,\parallel}(\theta^\prime,z_{2})+\psi_{\theta,\parallel}(\theta^\prime,z_{1})\psi_{\theta,\parallel}(\theta,z_{2})\right]\\ \nonumber
    & + U_{\theta, \parallel}(\boldsymbol{\mu}_{\parallel},\bold{r})U_{\theta, z}(\mu_{z},\bold{r}^\prime)e^{i(\alpha_{p}+\alpha_{p}^\prime)/2}\left[\psi_{\theta,\parallel}(\theta,z_{1})\psi_{\theta,z}(\theta^\prime,z_{2})+\psi_{\theta,z}(\theta^\prime,z_{1})\psi_{\theta,\parallel}(\theta,z_{2})\right]\\ \nonumber
    &+ U_{\theta, z}(\mu_{z},\bold{r})U_{\theta, \parallel}(\boldsymbol{\mu}_{\parallel},\bold{r}^\prime)e^{i(\alpha_{p}+\alpha_{p}^\prime)/2}\left[\psi_{\theta,z}(\theta,z_{1})\psi_{\theta,\parallel}(\theta^\prime,z_{2})+\psi_{\theta,\parallel}(\theta^\prime,z_{1})\psi_{\theta,z}(\theta,z_{2})\right]\\ 
    &+ U_{\theta, z}(\mu_{z},\bold{r})U_{\theta, z}(\mu_{z},\bold{r}^\prime)e^{i(\alpha_{p}+\alpha_{p}^\prime)/2}\left[\psi_{\theta,z}(\theta,z_{1})\psi_{\theta,z}(\theta^\prime,z_{2})+\psi_{\theta,z}(\theta^\prime,z_{1})\psi_{\theta,z}(\theta,z_{2})\right]=0.
\label{Psi polarizations}
\end{align}
Substituting Equations~\eqref{U}-\eqref{psi} into Equation~\eqref{Psi polarizations} it is straightforward to show that each of the four lines in the equation above has the same form as Equation~\eqref{wavefunction_substrate} and thus also admits substrate-independent solutions. Since the same  applies to $\Psi_{\theta,\phi}$, $\Psi_{\phi,\theta}$ and $\Psi_{\phi,\phi}$, the $\varepsilon$-independent zeros are also agnostic to the orientation of the QEs. However, it is important to note that the orientations of both QEs must be identical to preserve the indistinguishability of the emitted photons.

If we take the perfect mirror limit in Equation~\eqref{wavefunction_substrate}, the condition $\Psi=0$ reduces to
\begin{equation}
\cos(k_{0}z_{1}\cos\theta)\cos(k_{0}z_{2}\cos\theta^\prime)+\cos(k_{0}z_{1}\cos\theta^\prime)\cos(k_{0}z_{2}\cos\theta)=0.
\end{equation}
This expression indicates that when both QEs are positioned at $z_{1}, z_{2} < \lambda_{0}/4$, all the cosine functions take positive values, and thus $\Psi$ does not vanish. However, in contrast to the case of free space, there is no a minimum value of $z_{12}$ required for the emergence of geometrical zeros with a perfect mirror.

\subsection{Quantum emitter dimer in the vicinity of a nanosphere}

The far-field part of the scattered emission by a sphere of radius $R$ placed at the coordinate origin when a QE oriented along the $z$-axis is placed at $\bold{r}=b\hat{\boldsymbol{z}}$ (see the sketch in Figure~\ref{fig4}(a)) can be written as~\cite{buhmann}
\begin{equation}
    \hat{E}_{\theta}^{(+)}(\bold{r})=\frac{\omega_{0}^2\mu}{4\pi\epsilon_0c^2}\frac{e^{ik_{0}r}}{r}\sum_{l=1}^{\infty}c_{l}(\varepsilon,k_{0}R,k_{0}b)P_{l}^{1}(\cos\theta)\hat{\sigma},
\label{field_sphere}
\end{equation}
where $P_{l}^{1}(\cos\theta)$ are the associated Legendre polynomials and $c_{l}$ are the multipole coefficients defined as
\begin{equation}
    c_{l}=(-1)^{l}(2l+1)r_{l,\rm TM}(\varepsilon,k_{0}R,k_{0}b)\frac{h_{l}^{(1)}(k_{0}b)}{k_{0}b}i^{l+1}. \label{c}
\end{equation}
If the emitter is placed at $\bold{r}=-b\hat{\boldsymbol{z}}$, the multipole coefficients are
\begin{equation}
    \tilde{c}_{l}=-(2l+1)r_{l,\rm TM}(\varepsilon,k_{0}R,k_{0}b)\frac{h_{l}^{(1)}(k_{0}b)}{k_{0}b}i^{l+1}.\label{cvar}
\end{equation}
In Equations~\eqref{c} and~\eqref{cvar}, $r_{l,\rm TM}$ is the $l$-th order TM Mie reflection coefficient and $h_{l}^{(1)}(\cdot)$ the spherical Hankel functions of the first kind. Each term in the sum over $l$ in Equation~\eqref{field_sphere} corresponds to a different multipolar mode sustained by the sphere. The parameters $\gamma_{ii}$, $\gamma_{12}$ and $g_{12}$ in Equation~\eqref{master_eq} for this geometry are given by
\begin{equation}
    \gamma_{ii}=\gamma_{0}\left(1+\Im\left\{\frac{3i}{2}\sum_{l=1}^{\infty}(2l+1)l(l+1)r_{l,\rm TM}\left(\frac{h_{l}^{(1)}(k_{0}b)}{k_{0}b}\right)^{2}\right\}\right),
\end{equation}
\begin{equation}
    \gamma_{12}=\gamma_{0}\left(\frac{\gamma_{12,0}}{\gamma_{0}}+\Im\left\{\frac{3i}{2}\sum_{l=1}^{\infty}(2l+1)l(l+1)r_{l,\rm TM}(-1)^{l+1}\left(\frac{h_{l}^{(1)}(k_{0}b)}{k_{0}b}\right)^{2}\right\}\right),
\end{equation}
and
\begin{equation}
    g_{12}=\gamma_{0}\left(\frac{g_{12,0}}{\gamma_{0}}-\Re\left\{\frac{3i}{4}\sum_{l=1}^{\infty}(2l+1)l(l+1)r_{l,\rm TM}(-1)^{l+1}\left(\frac{h_{l}^{(1)}(k_{0}b)}{k_{0}b}\right)^{2}\right\}\right).
\end{equation}
In these expressions, $\gamma_{12,0}$ and $g_{12,0}$ are the parameters for the QE dimer in free space
\begin{equation}
    \gamma_{12,0}=3\gamma_{0}\frac{\sin(2k_{0}b)-2k_0b\cos(2k_0b)}{(2k_0b)^3},
\end{equation}
\begin{equation}
    g_{12,0}=-\frac{3}{2}\gamma_{0}\frac{\cos(2k_0b)+2k_0b\sin(2k_0b)}{(2k_{0}b)^3}.
\end{equation}

The condition for the occurrence of geometrical zeros can be written, using Equation~\eqref{field_sphere} and its counterpart in free space, as
\begin{align}\nonumber
   \Psi & =e^{-ik_{0}b\cos\theta}e^{ik_{0}b\cos\theta^\prime}+e^{-ik_{0}b\cos\theta^\prime}e^{ik_{0}b\cos\theta}\\ \nonumber
    &  + \sum_{l=1}^{\infty}f_{l}(\cos\theta^\prime)\left(\Tilde{c}_{l}e^{-ik_{0}b\cos\theta} + c_{l}e^{ik_{0}b\cos\theta}\right ) + \sum_{l=1}^{\infty}f_{l}(\cos\theta)\left(\Tilde{c}_{l}e^{-ik_{0}b\cos\theta^\prime} + c_{l}e^{ik_{0}b\cos\theta^\prime}\right ) \\ 
    & + \sum_{l,l^\prime=1}^{\infty}f_{l}(\cos\theta)f_{l^\prime}(\cos\theta^\prime)(c_{l}\tilde{c}_{l^\prime}+c_{l^\prime}\tilde{c}_{l})=0.
    \label{wavefunction_sphere}
\end{align}
where we define the functions $f_{l}(\cdot)$ as the first derivate of the Legendre polynomials and we use the property $P_{l}^{1}(\cos \theta)=-\sin \theta f_{l}(\cos \theta)$. In this case, the decomposition of Equation~\eqref{field_sphere} gives $U(\bold{r})=-\sin\theta(e^{ik_{0}r}/4\pi r)$ and $\psi(\theta, b\bold{\hat{z}})=\sum_{l=1}^{\infty}c_{l}(\varepsilon,k_{0}R,k_{0}b)f_{l}(\cos\theta)$ or $\psi(\theta, -b\bold{\hat{z}})=\sum_{l=1}^{\infty}\Tilde{c}_{l}(\varepsilon,k_{0}R,k_{0}b)f_{l}(\cos\theta)$, depending on the position of the QE.
Similar to the case of the flat substrate, there are four terms, each representing different parts of the two-photon optical path interference: the first term accounts for the interference associated with the two QE dipoles, the second and third terms account for the interference between the QE dipole and the nanosphere multipoles, and the fourth term accounts solely for the spherical multipoles. 

The shape of the geometrical-zero curves in the $\theta\theta^\prime$-plane varies depending on the dominant mutipolar mode sustained by the nanosphere. Importantly, this geometry does not admit $\varepsilon$-independent zeros. The reason can be traced back to the last term of the expression above. For this term to vanish, it must either be due to the cancellation of all function products $f_{l}(\cos\theta)f_{l^\prime}(\cos\theta^\prime)$, which would imply no far-field emission or it must be due to a specific choice of coefficients $c_{l}$ and $\tilde{c}_{l}$ that makes this term zero for two particular angles. However, since these coefficients depend on $\varepsilon$, they cannot be determined without specifying a particular value for the permittivity.

\begin{figure}[h!]
\centering
    \includegraphics[width=0.95\textwidth]{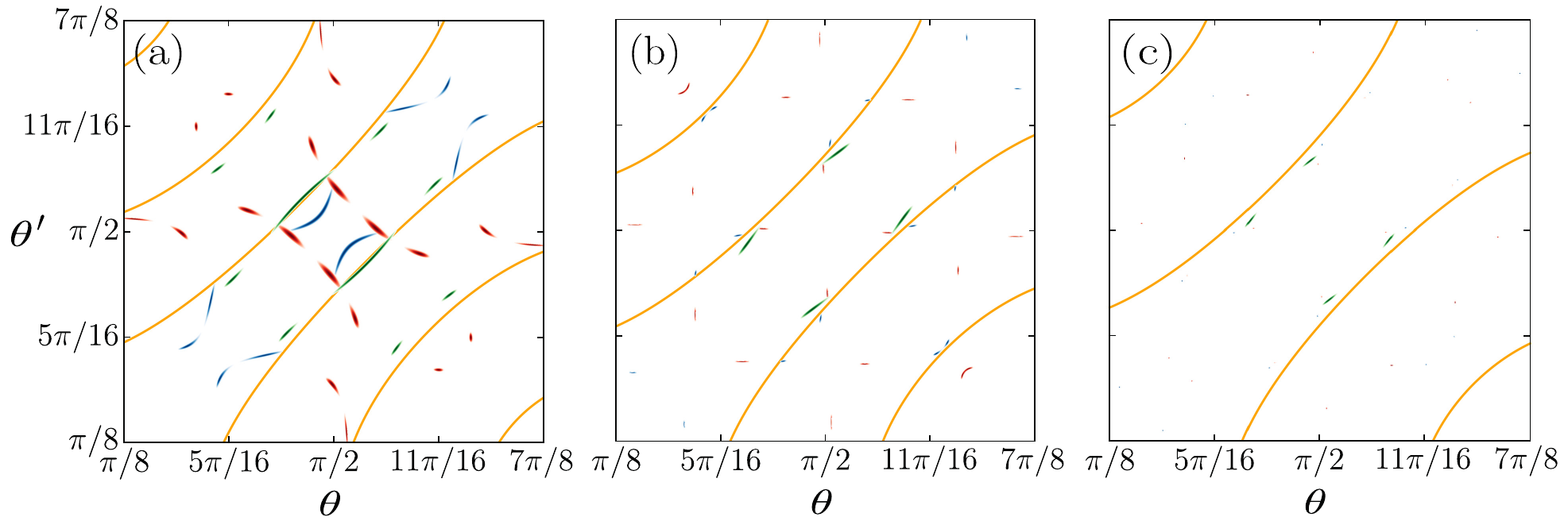}
    \caption{Map of the minima of $|\Psi|^{2}$ evaluated from Equation~\eqref{wavefunction_sphere} for different values of $z_{12}$: $1.45\lambda_{0}$ (a), $1.21\lambda_{0}$ (b), and $0.99\lambda_{0}$ (c). Similar to Figure~\ref{fig4}(c), blue, red, green, and orange curves correspond to $\varepsilon$: $-5+0.1i$, $-3+0.01i$,  $2.13$, and $1$, respectively.}
   \label{FIG_2_SM}
\end{figure}

Figure~\ref{FIG_2_SM} explores how the maps of the minima of $|\Psi|^{2}$ vary with the value of $z_{12}$. In all of the panels, we consider the same values of $\varepsilon$ as in Figure~\ref{fig4}(c). Analyzing these results, we observe that the shape of the curves signaling the minima of $|\Psi|^{2}$ strongly depends on $\varepsilon$ for all distances under consideration. However, as the QEs approach the sphere surface, their spatial extent is reduced.


%

\end{document}